\documentclass[pdflatex,sn-mathphys-num]{sn-jnl} 

\usepackage{amsmath,amsfonts}
\usepackage{array}
\usepackage{textcomp}
\usepackage{stfloats}
\usepackage{url}
\usepackage{verbatim}
\usepackage{graphicx}
\usepackage{tikz}
\usetikzlibrary{angles,quotes}
\usepackage{tikz-3dplot}
\usepackage{pgfplots}
\pgfplotsset{compat=1.18}
\usepackage{subcaption}
\usepackage{bm}
\usepackage{setspace}
\usepackage{geometry}
\DeclareMathOperator{\atantwo}{atan2}

\usepackage{physics}
\usepackage{fullwidth}
\usepackage{multirow}
\usepackage{xstring}
\newcommand{\mat}[1]{\mathbf{#1}}
\newcommand{\vv}[1]{\mathbf{#1}}

\def\Radius{2}

\newcommand{\plotvectortrigo}[5]{
    \def\x{{\Radius*cos(#3)*cos(90-#2)}}
    \def\y{{\Radius*cos(#3)*sin(90-#2)}}
    \def\z{{\Radius*sin(#3)}}
    \draw [->, -latex,#1,line width = 1.5pt] (0,0,0) -- ( \x,\y,\z)node[#4] {#5};
}
\newcommand{\plotvectorwithshadowtrigo}[5]{ 
    \def\x{{\Radius*cos(#3)*cos(90-#2)}}
    \def\y{{\Radius*cos(#3)*sin(90-#2)}}
    \def\z{{\Radius*sin(#3)}}
    \draw [->, -latex,#1,line width = 1.5pt] (0,0,0) -- (\x,\y,\z)node[#4] {#5};
    \draw[#1, dotted, thick] (0,0,0) -- (\x,\y,0);
    \draw[#1, dotted, thick] (\x,\y,0) -- (\x,\y,\z);
}

\newcommand{\defaultsphere}[3]{
    \path (0,0,0) coordinate (O);
    \def\Radius{#1}
    \pgfmathsetmacro{\cosangle}{cos(#3)}
    \pgfmathsetmacro{\sinangle}{sin(#3)}

    \shade[ball color=white,tdplot_screen_coords,opacity=0.5] (O) circle[radius=\Radius];
    \foreach \X/\Y in {xy/z,yz/x,zx/y}
    {\begin{scope}[canvas is \X\space plane at \Y=\Radius]
     \fill circle[radius=2pt];
    \end{scope}}
    \draw[->, -latex, thin, black] (0,0,0) -- (1.5*\Radius,0,0) node[pos=1.1, black] {$S_2$};
    \draw[->,, -latex, thin, black] (0,0,0) -- (0,1.5*\Radius,0) node[pos=1.2, black] {$S_1$};
    \draw[->, -latex, thin, black] (0,0,0) -- (0,0,1.5*\Radius) node[pos=1.1, black] {$S_3$};

    \begin{scope}[canvas is xy plane at z=0]
        \pgfmathsetmacro{\viewangle}{mod(360 + 0, 360)}
        \ifdim\viewangle pt < 180 pt
            \draw[dashed, gray] (-\Radius*\cosangle,\Radius*\sinangle) arc[start angle=150,end angle=330,radius=\Radius]; 
            \draw[solid, gray] (\Radius*\cosangle, -\Radius*\sinangle) arc[start angle=-30,end angle=150,radius=\Radius]; 
        \else
            \draw[solid, gray] (\Radius,0) arc[start angle=0,end angle=180,radius=\Radius]; 
            \draw[solid, gray] (-\Radius,0) arc[start angle=180,end angle=360,radius=\Radius]; 
        \fi
    \end{scope}
}

\tdplotsetmaincoords{110}{-30}
\newlength\pgfplotswidth

\pgfdeclarelayer{layer1}
\pgfdeclarelayer{layer2}
\pgfdeclarelayer{layer3}
\pgfdeclarelayer{layer4}
\pgfdeclarelayer{layer5}
\pgfdeclarelayer{layer6}
\pgfdeclarelayer{layer7}
\pgfdeclarelayer{layer8}

\pgfsetlayers{main, layer8, layer7, layer6, layer5, layer4, layer3, layer2, layer1}

\tikzset{>=latex}
\tikzset{axis/.style={black, very thick, ->}}
\tikzset{ef/.style={very thick, red}}
\tikzset{vec/.style={black, -{Latex[length=0.8mm]}}}
\tikzset{every text node part/.style={align=center}}

\newcommand{\waveplate}[3]{%
    \begin{scope}[canvas is xz plane at y=1.2]
        \draw[line join=round, thick, fill=black!40] (#1,-1.2) rectangle (#1+0.2,1.2);
    \end{scope}

    \begin{scope}[canvas is xy plane at z=1.2]
        \draw[line join=round, thick, fill=black!25](#1,-1.2) rectangle (#1+0.2,1.2);
    \end{scope}

    \begin{scope}[canvas is yz plane at x=#1]
        \draw[line join=round, thick, fill=black!10] (-1.2,-1.2) rectangle (1.2,1.2);
        \draw[line join=round, thick, fill=white] (0,0) coordinate (B) circle (0.8cm);
        \draw[line join=round, thick] (-{0.8*cos(#2)}, -{0.8*sin(#2)}) -- ({0.8*cos(#2)},{0.8*sin(#2)}) coordinate (A);
        \draw[line join=round, dashed, thick] (-0.8,0) -- (0.8,0) coordinate (C);
        \draw[line join=round] (0.24,0) arc (0:#2:0.24);
        \node at (0.45,0.25) {#3};
    \end{scope}
}

\begin{document}
\title{Waveplate-Based Transformation to Multiple Arbitrary Polarization States under a Common Rigid Rotation}

\author*[1]{\fnm{Joong-Seon} \sur{Choe}}\email{jschoe@etri.re.kr}

\affil*[1]{
	\orgdiv{
		Electronics and Telecommunications Research Institute}, \orgaddress{\street{218 Gajeong-ro, Yuseong-gu}, \city{Daejeon}, \postcode{34129}, \country{Republic of Korea}}
}

\abstract{
	A systematic method is presented for transforming a pair of initial polarization states (SoPs) into a corresponding target pair using a fixed configuration of two quarter-waveplates and one half-waveplate. The transformation is modeled as a rigid-body rotation in the three-dimensional Stokes space, allowing analytical derivation of waveplate angles that realize the desired mapping.
	While most existing analytical approaches assume linear or orthogonal states, the proposed method supports arbitrary target SoPs, including elliptical and non-orthogonal configurations in the Stokes space, provided they share a common rigid-body rotation with the inputs. Although the derivation is formulated for a single pair of SoPs, the method naturally extends to multiple SoPs under the same rotation condition.
	This enables deterministic and scalable polarization control for applications such as quantum key distribution, polarization-based signal processing, and integrated photonic systems.
}


\maketitle

\section{Introduction}
Polarization plays a central role in a wide range of optical systems, including quantum communication \cite{lim_demonstration_2023}, photonic integrated circuits (PICs) \cite{choe_silica_2018,ng_gigabit_2025}, optical communication \cite{tanPolarizationDivisionMultiplexing2024}, and precision metrology.
As these technologies rely increasingly on the polarization state of light, the ability to control and transform polarization with high precision has become a critical requirement.
In many applications, it is necessary not only to transform a single state of polarization (SoP) into another, but also to simultaneously align multiple SoPs to a set of desired targets.
This requirement is particularly important in quantum key distribution (QKD) protocols such as BB84, which rely on maintaining orthogonal or mutually unbiased polarization bases—such as horizontal/vertical (H/V) and diagonal/anti-diagonal (D/A)—with high fidelity \cite{bennet_quantum_1984}.

Waveplate-based polarization control is a well-established technique, and transforming a single SoP can be accomplished with just two properly oriented quarter-wave plates (QWPs) \cite{reddy_polarization_2016}. However, in applications like QKD, it is often necessary to align multiple SoPs simultaneously. Existing analytical methods are typically limited to specific polarization bases—such as HVDA. To the best of our knowledge, no general analytical method has been proposed for cases where the target SoPs include elliptical polarizations. This remains an active area of research: very recently, a systematic design of arbitrary polarization retarders and controllers built from QWP/HWP sequences was reported for a single SoP \cite{gevorgyan_arbitrary_2025}, underscoring that the generalization to the simultaneous, multiple-SoP case addressed here remains a timely and open problem.

This difficulty is not merely a lack of intuition but stems from two compounding factors. First, the polarization trajectory induced by a QWP on the Poincar\'{e} sphere is a nonlinear function of both the waveplate angle and the initial SoP \cite{salazar-ariza_trajectories_2018}, so inverting even a single waveplate setting from a desired output is already non-trivial. Second, aligning two SoPs simultaneously compounds this: since both SoPs undergo the identical rigid rotation, their required angles cannot be solved independently, and a rotation satisfying both constraints must instead be found jointly. Absent a systematic method, this joint problem has so far been handled either by restricting the target to special bases such as HVDA, as in prior work, or by numerical search over the three-parameter QQH configuration space, which offers no guarantee of convergence or of finding an exact solution. The method developed here removes both limitations at once: it yields a closed-form, non-iterative solution for arbitrary elliptical target pairs, together with the explicit necessary-and-sufficient existence condition of Eq.~(\ref{eq:existence}), which prior basis-restricted or numerical-search approaches do not supply.

While conventional waveplate-based methods are effective for aligning single SoP or a pair of SoPs belonging to standard bases such as HVDA, they do not generalize to cases where the desired output SoPs include arbitrarily oriented elliptical polarizations.

This challenge is particularly relevant in PIC implementations of QKD encoders and decoders. PICs offer advantages such as compactness, high integration, and alignment-free operation.
However, since all optical behavior is fixed during fabrication, even sub-wavelength deviations in optical path lengths—arising from minute variations in waveguide width or etch depth—can induce significant shifts in the output SoPs.
Such sensitivity makes consistent system performance heavily dependent on fabrication precision, which is difficult to ensure with high yield.
Although active tuning using components such as microheaters can compensate for these deviations, this work considers scenarios where the system operates passively, without dynamic feedback control.
In such cases, the goal is to apply a global rigid-body rotation to the SoPs resulting from fabrication, not to match predefined ideal bases, but to align them in a way that enables near-optimal system performance based on the actual polarization states available.
Previously available methods do not support analytical alignment to arbitrarily oriented elliptical SoPs, which may be required under such fabrication-induced variations.

In this paper, a mathematically rigorous and physically feasible approach is presented for transforming a pair of SoPs under the condition that a common rigid-body rotation exists in Stokes space. This transformation condition arises from the fact that any transformation performed by ideal waveplates must correspond to a unitary operation, which manifests as a rigid rotation in Stokes space.
Although the derivation is formulated for a pair of input and output SoPs, the method naturally extends to any number of polarization states, provided that they are all related by a common rigid-body rotation. In such cases, selecting any non-collinear pair from the set is sufficient to determine the transformation applicable to the entire group.

By extending the TRIAD algorithm which is originally developed in spacecraft attitude control, we derive a closed-form procedure for calculating the angles of two QWPs and one HWP arranged in a fixed sequence. Our method ensures deterministic realization of the desired transformation through analytical computation rather than iterative search.

As with other polarization transformation algorithms, our method applies only when the input and output SoP pairs are related through a unitary transformation; section~\ref{section:triad} shows that this reduces to the single, easily checked scalar condition of Eq.~(\ref{eq:existence}), which is necessary and sufficient.
While this study focuses on rigid-body transformations between polarization states, the analytical framework presented here may serve as a foundation for future optimization approaches, including those accommodating non-unitary effects or fabrication-induced imperfections in integrated photonic systems.

\section{Background and Theoretical Framework}\label{sec:background}
\subsection{Mueller Matrix Representation of Waveplates}
The change of an SoP by birefringent optical components can be represented as a rigid rotation.
Consequently, when two or more SoPs pass through the components, they maintain their relative position.
This implies that we only need to consider the transformation of two SoPs that are not collinear in the Stokes space.
To uniquely define a rigid rotation, the two SoPs must be linearly independent, i.e., not collinear. If SoPs are collinear or only one SoP is given, infinitely many combinations of waveplate rotation angles can generate the same target state, making the solution underdetermined \cite{wangTwoNovelPolarization2005}.

The Mueller matrix describing a general waveplate is as follows:
\begin{equation}
	\begin{aligned}
		       & \mat{M}(\theta, \Delta\phi)                                          \\
		=      & \begin{bmatrix}
			         1 & 0                     & 0                     & 0                \\
			         0 & C^2+S^2\cos\Delta\phi & CS(1-\cos\Delta\phi)  & -S\sin\Delta\phi \\
			         0 & CS(1-\cos\Delta\phi)  & S^2+C^2\cos\Delta\phi & C\sin\Delta\phi  \\
			         0 & S\sin\Delta\phi       & -C\sin\Delta\phi      & \cos\Delta\phi
		         \end{bmatrix} \\
		\equiv &
		\begin{bmatrix}
			1 & 0                                                     & 0 & 0 \\
			0 & \multicolumn{3}{c}{\multirow{3}{*}{\Large $\mat{R}$}}         \\
			0 &                                                               \\
			0 &
		\end{bmatrix}\!
	\end{aligned}\label{eq:waveplate_matrix}
\end{equation}
where $\Delta\phi$, $C$, and $S$ are the phase retardance, cos(2$\theta$) and sin(2$\theta$), respectively, with $\theta$ being the angle between the fast axis of the waveplate and the horizontal axis \cite{hecht_optics_2017}.
$\mat{R}$ is the submatrix of the Mueller matrix that contains information of retardance and waveplate angle, and can easily be confirmed to be a rotation matrix for reduced Stokes vector, composed of latter three components of Stokes vector.
In the analysis of polarization transformations induced by waveplates, it is sufficient to consider the reduced Stokes vector and rotation submatrix of the Mueller matrix, as can be seen from the first row and column of Eq.~(\ref{eq:waveplate_matrix}).

Throughout this paper, the term \textit{rigid rotation} refers to a unitary transformation in the reduced Stokes space $(S_1, S_2, S_3)$, corresponding to a proper rotation in three-dimensional real space. This is the mathematical manifestation of polarization transformations implemented by ideal birefringent waveplates.

From the Mueller matrix, the matrices describing QWP and HWP are obtained as in Eqs.~(\ref{eq:R_qwp}) and (\ref{eq:R_hwp}).
\begin{align}
	\mat{R}_{\rm QWP}(\theta)= &
	\begin{bmatrix}
		\cos^2 2\theta           & \cos 2\theta\sin 2\theta & -\sin 2\theta \\
		\cos 2\theta\sin 2\theta & \sin^2 2\theta           & \cos 2\theta  \\
		\sin 2\theta             & -\cos 2\theta            & 0
	\end{bmatrix}\label{eq:R_qwp} \\
	\mat{R}_{\rm HWP}(\theta)= &
	\begin{bmatrix}
		\cos 4\theta & \sin 4\theta  & 0  \\
		\sin 4\theta & -\cos 4\theta & 0  \\
		0            & 0             & -1
	\end{bmatrix}.\label{eq:R_hwp}
\end{align}

To transform multiple SoPs into desired SoPs, various configurations in view of waveplate sequence, such as QHQ, QQH, and HQQ, can be used, where Q and H represent QWP and HWP, respectively.

\subsection{TRIAD algorithm}\label{section:triad}

Before determining a method for polarization conversion using waveplates, the corresponding rotation transformation method must first be found. Given two initial ($\vv{a}_i$, $\vv{b}_i$) and target SoPs ($\vv{a}_f$, $\vv{b}_f$), the final goal is to implement a rotation transformation $\mat{R}_{\rm total}$ that satisfies the following equation using a combination of three waveplates.
\begin{equation}
	\mat{R}_{\rm total}
	\begin{bmatrix}
		\begin{array}{ccc}
			\vv{a}_i & \vv{b}_i
		\end{array}
	\end{bmatrix}
	=
	\begin{bmatrix}
		\begin{array}{ccc}
			\vv{a}_f & \vv{b}_f
		\end{array}
	\end{bmatrix}.\label{eq:connection}
\end{equation}
Noting that $\mat{R}_{\rm total}$ is a 3$\times$3 matrix, the SoP $\vv{c}_i$, which is orthogonal to the given two initial SoPs, is defined as follows
\begin{equation}
	\vv{c}_i := \frac{\vv{a}_i\times\vv{b}_i}{|\vv{a}_i\times\vv{b}_i|}\label{eq:def_c},\!
\end{equation}
and $\vv{c}_f$ is defined similarly.
To ensure a unique transformation, $\vv{a}_i$ and $\vv{b}_i$ must be non-collinear, which is always true when they serve as reference SoPs.
Then $\mat{R}_{\rm total}$ would give
\begin{equation}
	\mat{R}_{\rm total}
	=
	\begin{bmatrix}
		\begin{array}{ccc}
			\vv{a}_f & \vv{b}_f & \vv{c}_f
		\end{array}
	\end{bmatrix}
	\begin{bmatrix}
		\begin{array}{ccc}
			\vv{a}_i & \vv{b}_i & \vv{c}_i
		\end{array}
	\end{bmatrix}^{-1}.\label{eq:R_total_meaning}
\end{equation}
This formulation corresponds to the TRIAD algorithm, which determines the unique rotation that maps one linearly independent set of three vectors
$(\vv{a}_i, \vv{b}_i, \vv{c}_i)$ to another one $(\vv{a}_f, \vv{b}_f, \vv{c}_f)$.

\smallskip\noindent\textbf{Remark (existence condition).} A rotation $\mat{R}_{\rm total}\in SO(3)$ satisfying Eq.~(\ref{eq:connection}) exists if and only if
\begin{equation}
    \vv{a}_i\cdot\vv{b}_i = \vv{a}_f\cdot\vv{b}_f. \label{eq:existence}
\end{equation}
\emph{Necessity} follows because any $\mat{R}\in SO(3)$ preserves inner products: if $\mat{R}_{\rm total}\vv{a}_i=\vv{a}_f$ and $\mat{R}_{\rm total}\vv{b}_i=\vv{b}_f$, then $\vv{a}_i\cdot\vv{b}_i=(\mat{R}_{\rm total}\vv{a}_i)\cdot(\mat{R}_{\rm total}\vv{b}_i)=\vv{a}_f\cdot\vv{b}_f$.
\emph{Sufficiency}: writing $\mat{M}_i=[\vv{a}_i\ \vv{b}_i\ \vv{c}_i]$ and $\mat{M}_f=[\vv{a}_f\ \vv{b}_f\ \vv{c}_f]$, condition~(\ref{eq:existence}) together with the unit norms of all four SoPs and the orthogonality of $\vv{c}_i,\vv{c}_f$ to their respective pairs means $\mat{M}_i$ and $\mat{M}_f$ share the same Gram matrix, $\mat{M}_i^\top\mat{M}_i=\mat{M}_f^\top\mat{M}_f$. It follows directly that $\mat{R}_{\rm total}=\mat{M}_f\mat{M}_i^{-1}$ from Eq.~(\ref{eq:R_total_meaning}) satisfies $\mat{R}_{\rm total}^\top\mat{R}_{\rm total}=\mat{I}$, i.e., $\mat{R}_{\rm total}\in O(3)$; the right-handed convention used to define $\vv{c}_i,\vv{c}_f$ in Eq.~(\ref{eq:def_c}) further fixes $\det(\mat{R}_{\rm total})=+1$, so $\mat{R}_{\rm total}\in SO(3)$.
Consequently, an initial/target pair violating Eq.~(\ref{eq:existence}) cannot be connected by \emph{any} rotation, regardless of the waveplate configuration used to implement it --- this is a geometric constraint intrinsic to the transformation problem, not a limitation of the present method. For instance, $\vv{a}_i=\tfrac{1}{\sqrt2}(1,0,1)$, $\vv{b}_i=\tfrac{1}{\sqrt2}(0,1,1)$, $\vv{a}_f=\tfrac{1}{\sqrt2}(1,0,-1)$, $\vv{b}_f=\tfrac{1}{\sqrt2}(0,-1,1)$ gives $\vv{a}_i\cdot\vv{b}_i=0.5\neq-0.5=\vv{a}_f\cdot\vv{b}_f$, so no single rotation --- hence no combination of waveplates --- can realize this transformation.

While $\mat{R}_{\rm total}$ mathematically represents a rotation in Stokes space, realizing this transformation in a polarization optics system requires constructing it from physical components. In particular, it must be implemented through a sequence of three birefringent waveplates, each with a specific orientation.

\subsection{Standard Orientation for SoP Pair Transformation}\label{section:standard}

The QQH configuration—consisting of two QWPs and one HWP—has been used in free-space QKD experiments, utilizing two linear polarization bases: rectilinear and diagonal \cite{li_microsatellite-based_2024, yin_satellite-based_2017,roger_real-time_2023}.
This configuration takes advantages of two properties: (1) a QWP can transform any SoP into a linear one, and (2) an HWP can change the azimuth and reverse the handedness of an SoP \cite{tan_real-time_2024}.
This method is directly applicable when the target consists of two linear SoPs, with QKD employing the BB84 protocol serving as a representative application \cite{bennet_quantum_1984}.

To illustrate the principle, consider two initial SoPs $\vv{a}_i$ and $\vv{b}_i$ to be transformed into target linear SoPs $\vv{a}_f$ and $\vv{b}_f$—typically, $\vv{a}_f$ as horizontal polarization and $\vv{b}_f$ as a linear state offset from the diagonal.
The derived vectors $\vv{c}_i$ and $\vv{c}_f$ are defined by Eq.~(\ref{eq:def_c}).

The transformation proceeds as follows:
\begin{enumerate}
	\item Rotation of the first QWP (QWP1) to the angle $\theta_1$ converts $\vv{c}_i$ into a linear SoP $\vv{c}^\prime$.
	      $\theta_1$ is derived from Eq.~(\ref{eq:R_qwp}), as following equation:
	      \begin{equation}
		      \begin{aligned}
			      \begin{bmatrix}
				      \begin{array}{ccc}
					      \vv{a}^\prime & \vv{b}^\prime & \vv{c}^\prime
				      \end{array}
			      \end{bmatrix}
			       & =\mat{R}_{\rm QWP}(\theta_1)
			      \begin{bmatrix}
				      \begin{array}{ccc}
					      \vv{a}_i & \vv{b}_i & \vv{c}_i
				      \end{array}
			      \end{bmatrix}                                             \\
			       & \text{where} \quad \theta_1 = \frac{1}{2}\atantwo(c_{i2}, c_{i1}).
		      \end{aligned}\label{eq:qwp1}
	      \end{equation}
	      $c_{i1}$ and $c_{i2}$ denote components of $\vv{c}_i$ along $S_1$ and $S_2$ axes in Stokes space, respectively.
	      Note that $\atantwo$() function is used to account for the quadrant of SoP \cite{press_numerical_2007}.

	\item Rotation of the second QWP (QWP2) to $\theta_2$ makes $\vv{c}^\prime$ change into $-\hat{\vv{s}}_3$ corresponding to left circular polarization.
	      Physically, this means orienting the axis of QWP2 at an angle of -45$^\circ$ with respect to the polarization direction of $\vv{c}^\prime$.
	      \begin{equation}
		      \begin{aligned}
			      \begin{bmatrix}
				      \begin{array}{ccc}
					      \vv{a}^{\prime\prime} & \vv{b}^{\prime\prime} & \vv{c}^{\prime\prime}
				      \end{array}
			      \end{bmatrix}
			       & =\mat{R}_{\rm QWP}(\theta_2)
			      \begin{bmatrix}
				      \begin{array}{ccc}
					      \vv{a}^\prime & \vv{b}^\prime & \vv{c}^\prime
				      \end{array}
			      \end{bmatrix}                                                                    \\
			       & \text{where} \quad \theta_2 = \frac{3}{4}\pi+\frac{1}{2}\atantwo(c_2^\prime, c_1^\prime).
		      \end{aligned}\label{eq:qwp2}
	      \end{equation}

	\item Since $\vv{a}_i$ has been converted to the linear SoP $\vv{a}^{\prime\prime}$ as described in steps 2 and 3, rotate the HWP to $\theta_3$ to transform it into the SoP $\hat{\vv{s}}_1$, which represents horizontal polarization.
	      \begin{equation}
		      \begin{aligned}
			      \begin{bmatrix}
				      \begin{array}{ccc}
					      \vv{a}_f & \vv{b}_f & \vv{c}_f
				      \end{array}
			      \end{bmatrix}
			       & =\mat{R}_{\rm HWP}(\theta_3)
			      \begin{bmatrix}
				      \begin{array}{ccc}
					      \vv{a}^{\prime\prime} & \vv{b}^{\prime\prime} & \vv{c}^{\prime\prime}
				      \end{array}
			      \end{bmatrix}                                                                     \\
			       & \text{where} \quad \theta_3 = \frac{1}{4}\atantwo(a_2^{\prime\prime}, a_1^{\prime\prime}).
		      \end{aligned}\label{eq:hwp}\!
	      \end{equation}

	      Combining Eqs.~(\ref{eq:R_total_meaning}), (\ref{eq:qwp1}), (\ref{eq:qwp2}), and (\ref{eq:hwp}) yields
	      \begin{equation}
		      \mat{R}_{\rm total} =\mat{R}_{\rm HWP}(\theta_3)\mat{R}_{\rm QWP}(\theta_2)\mat{R}_{\rm QWP}(\theta_1).\label{eq:total}
	      \end{equation}
\end{enumerate}
Explicit calculation of $\mat{R}_{\rm total}$ is unnecessary for obtaining the waveplate angles in the standard orientation.
This method yields simple expressions for these angles by choosing $\vv{c}^\prime$ as linear, $\vv{c}^{\prime\prime}$ as circular, and $\vv{a}_f$ as horizontal polarization.
However, this simplicity comes at the cost of applicability: the technique is limited to transformations that achieve the standard orientation.
\def\rotationSphere{-100}
\def\radiusSphere{1.5cm}

\input{angles.dat}

\def\sphere_width{0.3\textwidth}
\begin{figure}[htbp]
	\centering
	\begin{subfigure}[b]{\sphere_width}
		\centering
		\tdplotsetmaincoords{110}{-30}
		\begin{tikzpicture}[tdplot_main_coords,thick]
			\defaultsphere{1.8}{default}{30} 
			\plotvectorwithshadowtrigo{red}{\psiLonA}{\psiLatA}{left}{$\vv{a}_i$}
			\plotvectorwithshadowtrigo{blue}{\psiLonB}{\psiLatB}{left}{$\vv{b}_i$}
			\plotvectorwithshadowtrigo{teal, dashed}{\psiLonC}{\psiLatC}{right}{$\vv{c}_i$}
		\end{tikzpicture}
		\caption{initial orientation}
	\end{subfigure}
	\hspace{1cm}
	\begin{subfigure}[b]{\sphere_width}
		\centering
		\tdplotsetmaincoords{110}{-30}
		\begin{tikzpicture}[tdplot_main_coords,thick]
			\defaultsphere{1.8}{default}{30} 
			\plotvectorwithshadowtrigo{red}{\psiLonAprime}{\psiLatAprime}{right}{$\vv{a}^\prime$}
			\plotvectorwithshadowtrigo{blue}{\psiLonBprime}{\psiLatBprime}{above}{$\vv{b}^\prime$}
			\plotvectorwithshadowtrigo{teal, dashed}{\psiLonCprime}{\psiLatCprime}{below}{$\vv{c}^\prime$}
		\end{tikzpicture}
		\caption{after QWP1}
	\end{subfigure}
	\vskip\baselineskip

	\begin{subfigure}[b]{\sphere_width}
		\centering
		\tdplotsetmaincoords{110}{-30}
		\begin{tikzpicture}[tdplot_main_coords,thick]
			\defaultsphere{1.8}{default}{30} 
			\plotvectortrigo{red}{\psiLonApprime}{\psiLatApprime}{above }{$\vv{a}^{\prime\prime}$}
			\plotvectortrigo{blue}{\psiLonBpprime}{\psiLatBpprime}{above }{$\vv{b}^{\prime\prime}$}
			\plotvectortrigo{teal, dashed}{\psiLonCpprime}{\psiLatCpprime}{above right}{$\vv{c}^{\prime\prime}$}
		\end{tikzpicture}
		\caption{after QWP2}
	\end{subfigure}
	\hspace{1cm}
	\begin{subfigure}[b]{\sphere_width}
		\centering
		\tdplotsetmaincoords{110}{-30}
		\begin{tikzpicture}[tdplot_main_coords,thick]
			\defaultsphere{1.8}{default}{30} 
			\plotvectortrigo{red}{\psiLonAf}{\psiLatAf}{above}{$\vv{a}_f$}
			\plotvectortrigo{blue}{\psiLonBf}{\psiLatBf}{above left}{$\vv{b}_f$}
			\plotvectortrigo{teal,dashed}{\psiLonCf}{\psiLatCf}{below right}{$\vv{c}_f$}
		\end{tikzpicture}
		\caption{after HWP}
	\end{subfigure}
	\caption{Transform sequence of two SoPs using QQH configuration.
		(a) Given two SoPs ($\vv{a}_i$, $\vv{b}_i$) and introduced SoP($\vv{c}_i$).
		(b) $\vv{c}_i$ is transformed to a linear SoP $\vv{c}^\prime$.
		(c) $\vv{c}^{\prime}$ is transformed to left circular SoP $\vv{c}^{\prime\prime}$.
		(d) $\vv{a}^{\prime\prime}$ is transformed to horizontal SoP $\vv{a}_f$ equivalent to $\hat{\vv{s}}_1$. $\vv{b}_f$ typically exhibits a deviation from the ideal $\hat{\vv{s}}_2$ direction since no orthogonality constraint is imposed on the SoP pair.}
	\label{fig:bloch_spheres}
\end{figure}

In this work, no assumption is made that $\vv{a}_i$ and $\vv{b}_i$ are orthogonal in the Stokes space. Since $\theta_3$ is determined with respect to $\vv{a}^{\prime\prime}$ as described in Eq.~(\ref{eq:hwp}), the resulting $\vv{b}_f$ generally corresponds to a linear polarization that is offset from the ideal diagonal direction, as illustrated in Fig.~\ref{fig:bloch_spheres}(d). This rotated configuration is hereafter referred to as the \textit{standard orientation}.

It is worth emphasizing that, although rotating a single polarization state to a target state involves two degrees of freedom — the choice of rotation axis and rotation angle — the simultaneous transformation of two distinct polarization states requires fully specifying the rigid-body rotation. Consequently, all three degrees of freedom associated with an SO(3) rotation must be determined to ensure that both input states are mapped accurately to their respective targets.

\section{Generalized Transformation to Arbitrary Orientations}
\subsection{Derivation of a Generalized Transformation for multiple SoPs}

To extend the method beyond the standard orientation, we introduce a general procedure for transforming a pair of input SoPs into an arbitrary pair of output SoPs that are related by a rigid-body rotation, using the QQH configuration. The key idea is to reuse the analytical results derived for the standard orientation by constructing a virtual SoP frame that is mapped to it through the same rotation.

Let $\mat{R}_{\rm total}$ be the rotation matrix that maps the given initial SoPs $\vv{a}_i$ and $\vv{b}_i$ to the target SoPs $\vv{a}_f$ and $\vv{b}_f$, as described in Eq.~(\ref{eq:connection}).
We define a set of three linearly independent virtual SoPs $\hat{\vv{q}}$, $\hat{\vv{u}}$, and $\hat{\vv{v}}$, such that:
\begin{equation}
	\mat{R}_{\rm total}
	\begin{bmatrix}
		\begin{array}{ccc}
			\hat{\vv{q}} & \hat{\vv{u}} & \hat{\vv{v}}
		\end{array}
	\end{bmatrix}
	=
	\begin{bmatrix}
		\begin{array}{ccc}
			\hat{\vv{s}}_1 & \hat{\vv{s}}_2 & \hat{\vv{s}}_3
		\end{array}
	\end{bmatrix},\label{eq:virtual_rotation}
\end{equation}
where $\hat{\vv{s}}_1$, $\hat{\vv{s}}_2$, and $\hat{\vv{s}}_3$ are the standard basis vectors in the Stokes space.
These virtual SoPs are not physically realized polarization states, but rather mathematical constructs introduced to allow the reuse of the known QQH angle solutions.

Since the canonical basis vectors $\hat{\vv{s}}_1$, $\hat{\vv{s}}_2$, and $\hat{\vv{s}}_3$ are orthonormal, matrix formed by them is the identity matrix:
\begin{equation}
	\begin{bmatrix}
		\begin{array}{ccc}
			\hat{\vv{s}}_1 & \hat{\vv{s}}_2 & \hat{\vv{s}}_3
		\end{array}
	\end{bmatrix}=\mat{I}.\label{eq:standard_basis}
\end{equation}
Since the right-hand side of Eq.~(\ref{eq:virtual_rotation}) is the identity matrix by Eq.~(\ref{eq:standard_basis}), the virtual SoPs are given by the inverse of the rotation matrix:
\begin{equation}
	\begin{bmatrix}
		\begin{array}{ccc}
			\hat{\vv{q}} & \hat{\vv{u}} & \hat{\vv{v}}
		\end{array}
	\end{bmatrix}
	= \mat{R}_{\rm total}^{-1}=\mat{R}_{\rm total}^\top.\label{eq:virtual_sop}
\end{equation}
The second equality follows from the fact that $\mat{R}_{\rm total}$ is a rotation matrix, whose inverse is equal to its transpose.
This compact result allows us to obtain the virtual SoPs directly from the transpose of the known rotation matrix without computing an explicit matrix inverse.

This step effectively rewrites the original problem--mapping $\vv{a}_i$ and $\vv{b}_i$ to $\vv{a}_f$ and $\vv{b}_f$--as an equivalent one where $\hat{\vv{q}}$ and $\hat{\vv{v}}$ are mapped to $\hat{\vv{s}}_1$ and $\hat{\vv{s}}_2$, respectively.
Since the waveplate angles required for this standard transformation have already been derived in Section~\ref{section:standard}, they can now be directly applied to the virtual SoPs $\hat{\vv{q}}$ and $\hat{\vv{u}}$.
This decouples the general SoP transformation problem into a coordinate rotation followed by a fixed optical transformation.

Specifically, by substituting $\hat{\vv{q}}$, $\hat{\vv{u}}$, and $\hat{\vv{v}}$ in place of $\vv{a}_i$, $\vv{b}_i$, and $\vv{c}_i$ in Eqs.~(\ref{eq:qwp1}), (\ref{eq:qwp2}), and (\ref{eq:hwp}), the final waveplate angles $\theta_1$, $\theta_2$, and $\theta_3$ required to implement $\mat{R}_{\rm total}$ are obtained.

This generalized approach enables deterministic calculation of waveplate settings for arbitrary target orientations, as long as a common rigid-body rotation exists between input and output SoPs. It significantly broadens the applicability of the method beyond special cases (e.g., linear bases) and avoids the need for iterative numerical optimization. Moreover, the method provides a compact and modular interpretation of polarization control in terms of rotation alignment and canonical transformation reuse.

\input{example_verify_dump.dat}

\subsection{Example}\label{sec:example}
In this section, we will explain an example of converting arbitrary SoPs, represented by any two independent Stokes vectors, into desired target SoPs.
Since this study deals with situations where SoPs undergo unitary transformations, the condition that must be satisfied is that the inner product between the two SoPs before and after the transformation remains preserved.
The assumed latitudes and longitudes of the vectors $\vv{a}_i$, $\vv{b}_i$, $\vv{a}_f$, and $\vv{b}_f$ are listed in Table~\ref{tab:lat_long_values}.
\begin{table}[htbp]
	\centering
	\begin{tabular}
		{c>{\centering\arraybackslash}p{1.2cm}>{\centering\arraybackslash}p{1.2cm}>{\centering\arraybackslash}p{1.2cm}>{\centering\arraybackslash}p{1.2cm}}
		\hline                     & \multicolumn{1}{c}{$\vv{a}_i$} & \multicolumn{1}{c}{$\vv{b}_i$} & \multicolumn{1}{c}{$\vv{a}_f$} & \multicolumn{1}{c}{$\vv{b}_f$} \\
		\hline Latitude ($^\circ$) & 60.00                          & -20.00                         & {\Latat}                       & {\Latbt}                       \\
		Longitude ($^\circ$)       & 22.50                          & -56.25                         & {\Lonat}                       & {\Lonbt}                       \\
		\hline
	\end{tabular} \caption{Latitude and longitude on the Poincar\'{e} sphere for $\vv{a}_i$, $\vv{b}_i$, $\vv{a}_f$, and $\vv{b}_f$} \label{tab:lat_long_values}
\end{table}

Using this information, we obtain each vector and can calculate the three coordinates of $\vv{c}_i$ and $\vv{c}_f$ to derive two 3$\times$3 matrices.
\begin{equation*}
	\begin{bmatrix}
		\begin{array}{ccc}
			\vv{a}_i & \vv{b}_i & \vv{c}_i\end{array}
	\end{bmatrix}
	=
	\begin{bmatrix}
		\begin{array}{ccc}
			{\Mizz} & {\Mizo} & {\Mizt} \\
			{\Mioz} & {\Mioo} & {\Miot} \\
			{\Mitz} & {\Mito} & {\Mitt}
		\end{array}
	\end{bmatrix}
\end{equation*}

\begin{equation*}
	\begin{bmatrix}
		\begin{array}{ccc}
			\vv{a}_f & \vv{b}_f & \vv{c}_f
		\end{array}
	\end{bmatrix}
	=
	\begin{bmatrix}
		\begin{array}{ccc}
			{\Mtzz} & {\Mtzo} & {\Mtzt} \\
			{\Mtoz} & {\Mtoo} & {\Mtot} \\
			{\Mttz} & {\Mtto} & {\Mttt}
		\end{array}
	\end{bmatrix}
\end{equation*}

Equation (\ref{eq:total}) is as follows:
\begin{equation*}
	\mat{R}_{\rm total}=
	\begin{bmatrix}
		\begin{array}{ccc}
			{\Mtotalzz} & {\Mtotalzo} & {\Mtotalzt} \\
			{\Mtotaloz} & {\Mtotaloo} & {\Mtotalot} \\
			{\Mtotaltz} & {\Mtotalto} & {\Mtotaltt}
		\end{array}
	\end{bmatrix}
\end{equation*}
Therefore, the three basis column vectors of the body frame, which become the three basis vectors of the reference coordinate system through the transformation, are as follows:
\begin{equation*}
	\begin{bmatrix}
		\begin{array}{ccc}
			\hat{\vv{q}} & \hat{\vv{u}} & \hat{\vv{v}}
		\end{array}
	\end{bmatrix}
	=
	\begin{bmatrix}
		\begin{array}{ccc}
			{\Mtotalzz} & {\Mtotaloz} & {\Mtotaltz} \\
			{\Mtotalzo} & {\Mtotaloo} & {\Mtotalto} \\
			{\Mtotalzt} & {\Mtotalot} & {\Mtotaltt}
		\end{array}
	\end{bmatrix}
\end{equation*}
The transformation of these three column vectors to standard orientation can be achieved by applying the method from section \ref{section:standard}, and the final results are as follows:
\begin{align*}
	\theta_1 & = {\firstqwpangle}^\circ  \\
	\theta_2 & = {\secondqwpangle}^\circ \\
	\theta_3 & = {\hwpangle}^\circ
\end{align*}
The procedure for determining $\hat{\vv{q}}$, $\hat{\vv{u}}$, and $\hat{\vv{v}}$ and then obtaining the waveplate rotation angles to transform $\vv{a}_i$, $\vv{b}_i$ to $\vv{a}_f$, $\vv{b}_f$ is depicted in Fig. \ref{fig:3pics}.
\begin{figure}[htbp]
	\centering
	\begin{subfigure}[b]{\sphere_width}
		\centering
		\tdplotsetmaincoords{110}{-30}
		\begin{tikzpicture}[tdplot_main_coords,thick]
			\defaultsphere{1.8}{default}{30}
			\plotvectorwithshadowtrigo{red}{\psiLonA}{\psiLatA}{left}{$\vv{a}_i$}
			\plotvectorwithshadowtrigo{red}{\psiLonB}{\psiLatB}{above left}{$\vv{b}_i$}
			\plotvectorwithshadowtrigo{blue}{\Lonat}{\Latat}{right}{$\vv{a}_f$}
			\plotvectorwithshadowtrigo{blue}{\Lonbt}{\Latbt}{above left}{$\vv{b}_f$}
		\end{tikzpicture}
		\caption{Obtain $\mat{R}_{\rm total}^\top$ from the given initial and target SoPs}
	\end{subfigure}
	\hspace{1cm}
	\begin{subfigure}[b]{0.2\textwidth}
		\centering
		\hspace{-3cm}
		\vspace{2.5cm}
		\begin{tikzpicture}
			\node (A) at (0,0) {};
			\node (B) at (1.5,1) {};
			\draw[->, scale=5, bend right ] (A) to (B) node[above] {\Large{$\mat{R}^\top_{\rm total}$}};
		\end{tikzpicture}
	\end{subfigure}
	\hspace{-1cm}
	\begin{subfigure}[b]{\sphere_width}
		\centering
		\tdplotsetmaincoords{110}{-30}
		\begin{tikzpicture}[tdplot_main_coords,thick, shift={(-8cm,0)}]
			\defaultsphere{1.8}{default}{30}
			\plotvectorwithshadowtrigo{red}{\psiLonA}{\psiLatA}{above}{$\vv{a}_i$}
			\plotvectorwithshadowtrigo{red}{\psiLonB}{\psiLatB}{above left}{$\vv{b}_i$}
			\plotvectorwithshadowtrigo{teal, dashed}{\Lonq}{\Latq}{above left}{$\hat{\vv{q}}$}
			\plotvectorwithshadowtrigo{teal, dashed}{\Lonu}{\Latu}{above left}{$\hat{\vv{u}}$}
			\plotvectorwithshadowtrigo{teal, dashed}{\Lonv}{\Latv}{right}{$\hat{\vv{v}}$}
		\end{tikzpicture}
		\caption{Determination of $\hat{\vv{q}}$, $\hat{\vv{u}}$, and $\hat{\vv{v}}$ SoPs from column vectors of $\mat{R}_{\rm total}^\top$}
	\end{subfigure}
	\vskip\baselineskip

	\begin{subfigure}[b]{0.45\textwidth}
		\centering
		\tdplotsetmaincoords{110}{-30}
		\begin{tikzpicture}[x={(1cm,0.0cm)}, y={(8mm, -3mm)}, z={(0cm,1cm)}, line cap=round, line join=round, shift={(-3cm, 0cm)}]
			\raisebox{0.cm}{
				\draw[axis] (4,0,0) -- (6,0,0);

				\begin{pgfonlayer}{layer1}
					\waveplate{1}{60}{$\theta_1$}
				\end{pgfonlayer}
				\begin{pgfonlayer}{layer3}
					\waveplate{2.5}{144}{$\theta_2$}
				\end{pgfonlayer}
				\begin{pgfonlayer}{layer8}
					\waveplate{4}{105}{$\theta_3$}
				\end{pgfonlayer}

				\begin{scope}[canvas is yz plane at x=0.5]
					\node[rotate=-20] at (0.5,1.8) {\small{QWP1}};
				\end{scope}

				\begin{scope}[canvas is yz plane at x=2.]
					\node[rotate=-20] at (0.5,1.8) {\small{QWP2}};
				\end{scope}

				\begin{scope}[canvas is yz plane at x=3.5]
					\node[rotate=-20] at (0.5,1.8) {\small{HWP}};
				\end{scope}

				\begin{pgfonlayer}{layer1}
					\draw[very thick] (-0.3,0,0) -- (1,0,0);
				\end{pgfonlayer}
				\begin{pgfonlayer}{layer3}
					\draw[very thick] (2,0,0) -- (2.5,0,0);
				\end{pgfonlayer}
				\begin{pgfonlayer}{layer5}
					\draw[line cap=round, very thick] (3,0,0) -- (4,0,0);
				\end{pgfonlayer}
			}
		\end{tikzpicture}
		\caption{QQH angles determination for transforming $\hat{\vv{q}}$ and $\hat{\vv{u}}$ into standard orientation}
	\end{subfigure}
	\hspace{2cm}
	\begin{subfigure}[b]{\sphere_width}
		\centering
		\tdplotsetmaincoords{110}{-30}
		\begin{tikzpicture}[tdplot_main_coords,thick]
			\defaultsphere{1.8}{empty}{30}
			\plotvectorwithshadowtrigo{blue}{\Lonat}{\Latat}{right}{$\vv{a}_f$}
			\plotvectorwithshadowtrigo{blue}{\Lonbt}{\Latbt}{above}{$\vv{b}_f$}
			\plotvectortrigo{teal, dashed}{0}{0}{above}{$\hat{\vv{s}}_1$}
			\plotvectortrigo{teal, dashed}{90}{0}{above}{$\hat{\vv{s}}_2$}
			\plotvectortrigo{teal, dashed}{90}{90}{left}{$\hat{\vv{s}}_3$}
		\end{tikzpicture}\caption{Target SoPs achieved by transforming $\hat{\vv{q}}$ and $\hat{\vv{u}}$ into standard orientation}
	\end{subfigure}
	\caption{Transformation procedure for mapping an initial SoP pair to an arbitrary target pair via rigid rotation.
		(a) Initial and target SoPs define a rotation $\mat{R}_{\rm total}$.(b) Virtual SoPs are constructed to map to the canonical basis under $\mat{R}_{\rm total}$.
		(c) The virtual SoPs undergo waveplate-based rotation using analytically derived angles.
		(d) The transformed SoPs align with their targets, and the virtual frame aligns with the standard axes.}
	\label{fig:3pics}
\end{figure}

\section{Discussion}\label{sec:discussion}
The analytical framework developed in this work enables deterministic calculation of waveplate angles for transforming a pair of SoPs and, by extension, a set of multiple SoPs, as long as a common rigid-body rotation exists between the initial and target SoPs. This condition is naturally satisfied in systems exhibiting unitary polarization behavior, such as lossless, reciprocal optical channels or well-aligned photonic modules. Unlike existing approaches that focus on canonical polarization bases or rely on numerical optimization, the proposed method generalizes to arbitrary target states—including elliptical SoPs—within the constraint of a shared unitary transformation.

This capability is especially relevant in integrated photonic systems, where fabrication-induced deviations in waveguide dimensions or optical path lengths often prevent perfect orthogonality among output channels. In such environments, rigid alignment to ideal bases like HV or DA may not yield optimal performance. Instead, polarization transformation toward non-standard—but system-specific optimal—SoP configurations can improve fidelity in polarization-encoded applications such as QKD. The present framework thus provides a mathematically grounded approach to align polarization states under non-ideal but realistic hardware conditions.

Additionally, the method's closed-form nature enables efficient implementation in calibration routines, feedback control systems, or PIC-based polarization encoders where low-latency operation and deterministic behavior are critical. By separating the geometric alignment (rigid rotation) from the optical implementation (waveplate angles), the technique also offers modularity and clarity in polarization system design.

A practical consideration for implementing the derived waveplate settings is the sensitivity of the transformation to alignment and retardance errors: because the evolution of a polarization state on the Poincar\'{e} sphere is highly sensitive to small deviations in fast-axis orientation and retardance, maintaining all three waveplates precisely on-axis and at their nominal quarter-/half-wave retardance is essential to realizing the target SoPs in practice. Mechanical mounts with sub-degree angular resolution, temperature-stabilized retarders, or an iterative fine-tuning step guided by polarimetric feedback can mitigate this sensitivity; a full experimental characterization of these tolerances is left for future work.

Future work will explore the extension of this framework to cases where a single rigid rotation does not suffice, such as non-unitary channels or scenarios requiring local transformations among multiple independent SoP pairs. Such generalizations may involve decompositions into piecewise rotations or optimization-guided corrections beyond the unitary domain, enabling more comprehensive polarization control across complex photonic platforms.

\section{Conclusion}\label{sec:conclusion}
In this study, a generalized mathematical framework was developed for transforming a pair of polarization states into an arbitrary target pair through rigid rotation. The method naturally extends to multiple polarization states by selecting any non-collinear initial and target pairs from the set, provided a common rigid-body rotation exists.
This approach, applicable to advanced photonic technologies such as quantum communication, optical sensing, and photonic signal processing, offers a robust solution for precise and adaptable polarization control.
Unlike conventional methods restricted to linear SoPs, our framework enables direct and efficient transformations between arbitrary polarization states, significantly enhancing the flexibility and accuracy of polarization manipulation.
By overcoming the limitations of existing approaches, this method provides a key stepping stone toward improved performance in real-world applications.
Future research will extend this framework beyond the single shared-rotation setting, addressing non-unitary channels and independent multi-pair transformations to further broaden its applicability to complex photonic platforms.

\bibliography{jschoe}

\section*{Funding}
This work was supported by the Institute of Information \& communications Technology Planning \& Evaluation (IITP) grant funded by the Korea government (MSIT) (2019-0-00005, Technology development of transmitter and receiver integrated module in a polarization based free-space quantum key distribution for short-range low-speed moving quantum communication), (RS-2026-25522225, Integrated Fiber-FSO Quantum-Secured Communication Technology for MUM-T)
\end{document}